\documentclass[twocolumn,showpacs,superscriptaddress,amsmath,amssymb]{revtex4}
\usepackage{graphicx}
\usepackage{dcolumn}
\usepackage{bm}

\def\bd{
\begin{document}} \def\ed{\end{document}}
\def\bmp{\begin{minipage}} \def\emp{\end{minipage}}
\def\bcc{\begin{center}} \def\ecc{\end{center}}     \def\npg{\newpage}
\def\beq{\begin{equation}} \def\eeq{\end{equation}} \def\hph{\hphantom}
\def\be{\begin{equation}} \def\ee{\end{equation}} \def\r#1{$^{[#1]}$}
\def\n{\noindent} \def\ni{\noindent} \def\pa{\parindent}
\def\hs{\hskip} \def\vs{\vskip} \def\hf{\hfill} \def\ej{\vfill\eject}
\def\cl{\centerline} \def\ob{\obeylines}  \def\ls{\leftskip}
\def\underbar#1{$\setbox0=\hbox{#1} \dp0=1.5pt \mathsurround=0pt
   \underline{\box0}$}   \def\ub{\underbar}    \def\ul{\underline}
\def\f{\left} \def\g{\right} \def\e{{\rm e}} \def\o{\over} \def\d{{\rm d}}
\def\vf{\varphi} \def\pl{\partial} \def\cov{{\rm cov}} \def\ch{{\rm ch}}
\def\la{\langle} \def\ra{\rangle} \def\EE{e$^+$e$^-$} \def\pt{p_{\rm t}}
\def\bitz{\begin{itemize}} \def\eitz{\end{itemize}}
\def\btbl{\begin{tabular}} \def\etbl{\end{tabular}}
\def\btbb{\begin{tabbing}} \def\etbb{\end{tabbing}}
\def\beqar{\begin{eqnarray}} \def\eeqar{\end{eqnarray}}
\def\\{\hfill\break} \def\dit{\item{-}} \def\i{\item}
\def\bbb{\begin{thebibliography}{9} \itemsep=-1mm}
\def\ebb{\end{thebibliography}} \def\bb{\bibitem}
\def\bpic{\begin{picture}(260,240)} \def\epic{\end{picture}}
\def\akgt{\cl{\bf ACKNOWLEDGMENTS}}
\def\fgn{\noindent{\bf\large\bf figure captions}}
\def\lan{\langle}
\def\ran{\rangle}
\def\p{\pi}
\def\ifmath#1{\relax\ifmmode #1\else $#1$\fi}%
\def\rc{\ifmath{{\mathrm{c}}}}
\def\cut{\ifmath{{\mathrm{cut}}}}
\def\rF{\ifmath{{\mathrm{F}}}}
\def\rK{\ifmath{{\mathrm{K}}}}
\def\rp{\ifmath{{\mathrm{p}}}}
\def\rt{\ifmath{{\mathrm{t}}}}
\def\LAB{\ifmath{{\mathrm{LAB}}}}
\def\cut{\ifmath{{\mathrm{cut}}}}
\def\beq{\begin{equation}}
\def\eeq{\end{equation}}

\newcommand{\cinst}[2]{$^{\mathrm{#1}}$~#2\par}
\newcommand{\crefi}[1]{$^{\mathrm{#1}}$}
\newcommand{\crefii}[2]{$^{\mathrm{#1,#2}}$}
\newcommand{\crefiii}[3]{$^{\mathrm{#1,#2,#3}}$}
\newcommand{\HRule}{\rule{0.5\linewidth}{0.5mm}}

\bd

\title{\boldmath Observation of a $p\bar{p}$  mass threshold enhancement
in \\
$\psi^\prime\rightarrow\pi^{+}\pi^{-}J/\psi(J/\psi\rightarrow\gamma
p\bar{p})$ decay}


\author{
M.~Ablikim$^{1}$, M.~N.~Achasov$^{5}$, L.~An$^{9}$, Q.~An$^{31}$,
Z.~H.~An$^{1}$, J.~Z.~Bai$^{1}$, Y.~Ban$^{18}$, N.~Berger$^{1}$,
J.~M.~Bian$^{1}$, I.~Boyko$^{13}$
% ...
\\
\vspace{0.2cm}
(BESIII Collaboration)\\
\vspace{0.2cm} {\it
$^{1}$ Institute of High Energy Physics, Beijing 100049, P. R. China\\
% ...
$^{5}$ G.I. Budker Institute of Nuclear Physics SB RAS (BINP), Novosibirsk 630090, Russia\\
% ...
$^{9}$ Guangxi University, Naning 530004, P. R. China\\
% ...
$^{13}$ Joint Institute for Nuclear Research, 141980 Dubna, Russia\\
% ...
$^{18}$ Peking University, Beijing 100871, P. R. China\\
% ...
$^{31}$ University of Science and Technology of China, Hefei 230026, P. R. China\\
% ...
}}

\vspace{0.4cm}

\date{\today}

\begin{abstract}

  The decay channel
  $\psi^\prime\rightarrow\pi^+\pi^-J/\psi(J/\psi\rightarrow\gamma
  p\bar{p})$ is studied using a sample of $1.06\times 10^8$ $\psi^\prime$
  events collected by the BESIII experiment at BEPCII. A strong
  enhancement at threshold is observed in the $p\bar{p}$ invariant
  mass spectrum. The enhancement can be fit with an $S$-wave
  Breit-Wigner resonance function with a resulting peak mass of
  $M=1861^{+6}_{-13}~{\rm (stat)}^{+7}_{-26}~{\rm (syst)}~{\rm
    MeV/}c^2$ and a narrow width that is $\Gamma<38~{\rm MeV/}c^2$ at
  the $90\%$ confidence level. These results are consistent with
  published BESII results.  These mass and width values do not match
  with those of any known meson resonance.

\end{abstract}


\maketitle An anomalously  strong $p\bar{p}$ mass threshold
enhancement was observed by the BESII experiment in the radiative
decay process $J/\psi\rightarrow\gamma p \bar{p}$~\cite{ppb_jixb}.
In ref.~\cite{ppb_jixb} it was noted that when an $S$-wave
Breit-Wigner resonance function is fitted to the  $p\bar{p}$ mass
distribution, the peak mass is below the $p\bar{p}$ mass threshold
at $M=1859^{+3}_{-10}~{\rm(stat)}^{+5}_{-25}~{\rm(syst)}~{\rm
MeV}/c^2$ and the total width  is $\Gamma<30~{\rm MeV}/c^2$ (at
the $90\%$ C.L.). An interesting feature of this enhancement is
that corresponding structures are not observed in near-threshold
$p\bar{p}$ cross section measurements,  in $B$-meson
decays~\cite{Bdecay,ichep04},  in radiative $\psi^\prime$ or
$\Upsilon\rightarrow\gamma p\bar{p}$ decays~\cite{psip,upslon}, or
in $J/\psi\rightarrow\omega p \bar{p}$ decays~\cite{omegappb}.
These non-observations disfavor the attribution of the
mass-threshold enhancement, which is uniquely observed in the
$J/\psi\rightarrow\gamma p\bar{p}$ decay process, to the pure
effects of $p\bar{p}$ final state interactions (FSI).

This experimental observation stimulated a number of theoretical
speculations~\cite{theory1,theory2,theory3,theory4,theory5,theory6}.
One of these is the intriguing suggestion that it is an example of a
$p\bar{p}$ bound state, sometimes called baryonium~\cite{baryonium},
which has a long history and has been the subject of many experimental
searches~\cite{eexp1}.

In this letter we report a study of the $p\bar{p}$ mass spectrum
in the threshold region in the decay process
$\psi^\prime\rightarrow\pi^{+}\pi^{-}J/\psi(J/\psi\rightarrow\gamma
p\bar{p})$.  The analysis uses a sample of $1.06 \times 10^{8}$
$\psi^\prime$ events accumulated by the upgraded Beijing
Spectrometer (BESIII) located at the Beijing Electron-Positron
Collider (BEPCII) at the Beijing Institute of High Energy Physics.

BEPCII/BESIII~\cite{bes3} is a major upgrade of the BESII
experiment at the BEPC accelerator~\cite{bes2}. The design peak
luminosity of the double-ring $e^+e^-$ collider, BEPCII, is
$10^{33}$ cm$^{-2}s^{-1}$ at a beam current of 0.93~A. The BESIII
detector with a geometrical acceptance of 93\% of 4$\pi$, consists
of the following main components: 1) a small-celled, helium-based
main draft chamber (MDC) with 43 layers.
The average single wire resolution is 135 $\mu m$, and the
momentum resolution for 1~GeV charged particles in a 1 T magnetic
field is 0.5\%; 2) an electromagnetic calorimeter (EMC) made of
6240 CsI (Tl) crystals arranged in a cylindrical shape (barrel)
plus two endcaps. For 1.0~GeV photons, the energy resolution is
2.5\% in the barrel and 5\% in the endcaps, and the position
resolution is 6~mm in the barrel and 9~mm in the endcaps; 3) a
Time-Of-Flight system (TOF) for particle identification composed
of a barrel part made of two layers with 88 pieces of 5~cm thick,
2.4~m long plastic scintillators in each layer, and two endcaps
with 96 fan-shaped, 5~cm thick, plastic scintillators in each
endcap.  The time resolution is 80~ps in the barrel, and 110 ps in
the endcaps, corresponding to a $2\sigma$ K/$\pi$ separation for
momenta up to about 1.0~GeV; 4) a muon chamber system (MUC) made
of 1000~m$^2$ of Resistive Plate Chambers (RPC) arranged in 9
layers in the barrel and 8 layers in the endcaps and incorporated
in the return iron of the superconducting magnet.  The position
resolution is about 2~cm.


Candidate $\psi^\prime\rightarrow\pi^+\pi^- J/\psi
(J/\psi\rightarrow\gamma p\bar{p})$ events are required to have at
least one photon and four charged tracks within the polar angle
region $|\cos\theta|<0.93$ and a total net charge of zero. The TOF
and $dE/dx$ information are combined to form particle
identification confidence levels for the $\pi$, $K$, and $p$
hypotheses; each track is assigned to the particle type that
corresponds to the hypothesis with the highest confidence level.
Selected events are required to have both an identified proton and
an identified anti-proton and no particle identification is
required for the two remaining tracks. Candidate photons are
required to have an energy deposit that is at least 25~MeV in the
barrel EMC $(|\cos\theta|<0.8)$) and 50~MeV in the endcap EMCs
$(0.86<|\cos\theta|<0.92)$,  and be isolated from the anti-proton
track by more than $30^{\circ}$ due to the strong annihilation of
anti-protons, and from all other charged tracks by more than
$10^{\circ}$. EMC timing requirements suppress electronic noise
and energy deposits unrelated to the event.

Candidate $J/\psi$ signals are identified by the invariant mass
recoiling against the $\pi^+\pi^-$ pair,
$|M_{\pi^+\pi^-_{recoil}}-m_{J/\psi}|<0.006~{\rm GeV/}c^2$.
Further requirements of $|U_{miss}|<0.05~{\rm GeV}$, where
$U_{miss}=(E_{miss}-|P_{miss}|)$,
 and $P^{2}_{t\gamma}<0.0005~({\rm GeV/}c)^2$, where
$P^{2}_{t\gamma}=4|P_{miss}|^2\sin^{2}\theta_{\gamma}/2$, are
imposed to suppress backgrounds from multi-photon events. Here
$E_{miss}$ and $P_{miss}$ are, respectively, the missing energy
and momentum of all charged particles, and  $\theta_{\gamma}$ is
the angle between the missing momentum and the photon direction.
The requirement
$|M_{\pi^{+}\pi^{-}p\bar{p}}-m_{\psi^\prime}|>0.03~{\rm GeV/}c^2$
is used to reduce the background from
$\psi^\prime\rightarrow\pi^{+}\pi^{-}p\bar{p}$.

Events that remain after these selection requirements are
subjected to a four-constraint energy-momentum conservation
kinematic fit to the hypothesis
$\psi^\prime\rightarrow\gamma\pi^{+}\pi^{-}p\bar{p}$. For events
with more than one $\gamma$ candidate, the combination with the
smallest $\chi^{2}$ is chosen.  Events with $\chi^{2}<100$ are
selected. Since the detection efficiencies for data and Monte
Carlo (MC) simulated events are consistent for protons and
antiprotons with momenta $p>0.3~{\rm GeV/}c$, while differences
occur for lower momentum tracks, we reject events with
$p_p<0.3~{\rm GeV/}c$ or $p_{\bar{p}}<0.3~{\rm GeV/}c$.

Fig.~\ref{Fig1}(a) shows the $p\bar{p}$ invariant mass
distribution for surviving events. The distribution features a
peak near $M_{p\bar{p}}=2.98~{\rm GeV/}c^2$ that is consistent in
mass, width, and yield with expectations for
$\psi^\prime\rightarrow\pi^{+}\pi^{-}J/\psi
(J/\psi\rightarrow\gamma\eta_{c}, \eta_c\rightarrow p\bar{p})$, a
broad enhancement around $M_{p\bar{p}}\sim 2.2~{\rm GeV/}c^2$, and
a prominent low-mass peak at the $p\bar{p}$ mass threshold,
similar to that reported by BESII~\cite{ppb_jixb}. The Dalitz plot
for selected events is shown in Fig.~\ref{Fig1}(b), where a band
corresponding to the threshold enhancement is evident in the upper
right corner.

\begin{figure}
\includegraphics[width=3.5in]{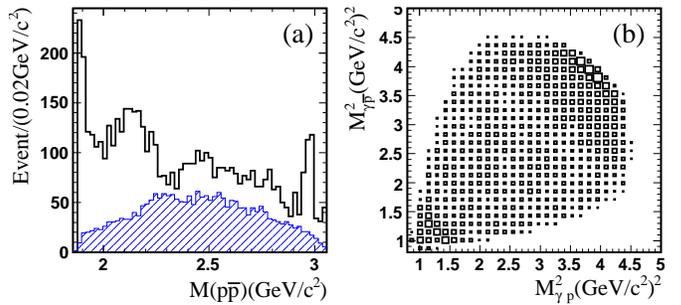}
\caption{\label{Fig1} The $p\bar{p}$ invariant mass spectrum for
the selected $\psi^\prime\rightarrow\pi^{+}\pi^{-}J/\psi
(J/\psi\rightarrow\gamma p\bar{p})$ candidate events. (a) The
$p\bar{p}$ invariant mass spectrum; the open histogram is data and
the hatched histogram is from a
$\psi^\prime\rightarrow\pi^{+}\pi^{-}J/\psi
(J/\psi\rightarrow\gamma p\bar{p})$ phase-space MC events(with
arbitrary normalization). (b) An $M^2(\gamma p)$~(horizontal)
$versus$ $M^2(\gamma \bar{p})$~(vertical) Dalitz plot for the
selected events.}
\end{figure}

Potential background processes are studied with an inclusive MC
sample of $1\times 10^{8}$ $\psi^\prime$ events generated
according to the Lund-Charm model~\cite{lund} and the Particle
Data Group (PDG) decay tables ~\cite{PDG}.  None of background
sources produce an enhancement at the threshold region of
$p\bar{p}$ invariant-mass spectrum.  The dominant background is
from
$\psi^\prime\rightarrow\pi^{+}\pi^{-}J/\psi(J/\psi\rightarrow\pi^{0}
p\bar{p})$ events, with asymmetric
$\pi^{0}\rightarrow\gamma\gamma$ decays where one of the photons
has most of the $\pi^{0}$ energy. An exclusive MC sample of
$\psi^\prime\rightarrow\pi^{+}\pi^{-}J/\psi(J/\psi\rightarrow\pi^{0}
p\bar{p})$, generated with a uniform phase space distribution,
indicates that the level of this background in the selected event
sample with $M_{p\bar{p}}-2m_{p}<0.3~{\rm GeV}/c^2$ is $9\%$ of
the total.

To ensure further that the $p\bar{p}$ threshold enhancement is not
due to background, each potential background  is studied with
data. Non-$J/\psi$  background are studied using $J/\psi$
mass-sideband events. For these there is no enhancement and their
level of contamination of the selected event sample is about
$2\%$. The dominant background channel,
 $\psi^\prime\rightarrow\pi^{+}\pi^{-}J/\psi(J/\psi\rightarrow\pi^{0}
p\bar{p})$, is also studied with data. In this case, events with
four charged tracks, including a proton and antiproton and two
oppositely charged pions, and with two or more photons are
selected, and subjected to a four-constraint kinematic fit to the
$\psi^\prime\rightarrow\gamma\gamma\pi^{+}\pi^{-}p\bar{p}$
hypothesis. $J/\psi$ and $\pi^0$ signals are selected by the
requirements $|M_{\pi^+\pi^-_{recoiling}}-m_{J/\psi}|<0.006~{\rm
GeV/}c^2$ and
 $|M_{\gamma\gamma}-m_{\pi^0}|<0.008~{\rm GeV/}c^2~(\pm 2\sigma)$.
There is no evidence of a narrow and strong enhancement near
$p\bar{p}$ mass threshold region.

The $M_{p\bar{p}}$ invariant mass spectrum in the threshold region
for the selected $\pi^0 p\bar{p}$ events is shown in
Fig.~\ref{Fig2}$(a)$,  where no threshold enhancement is evident.
The distribution is well described by a function of the form
$f_{bkg}(\delta)=N(\delta^{1/2}+a_{1}\delta^{3/2}+a_{2}\delta^{5/2})$,
where $\delta=M_{p\bar{p}}-2m_{p}$ and the shape parameters $a_1$
and $a_2$ are determined from a fit to selected $\gamma p\bar{p}$
events for $\psi^\prime\rightarrow\pi^{+}\pi^{-}J/\psi
(J/\psi\rightarrow\gamma p\bar{p})$ phase-space MC sample shown in
Fig.~\ref{Fig2}$(b)$.

\begin{figure}
\includegraphics[width=1.68in]{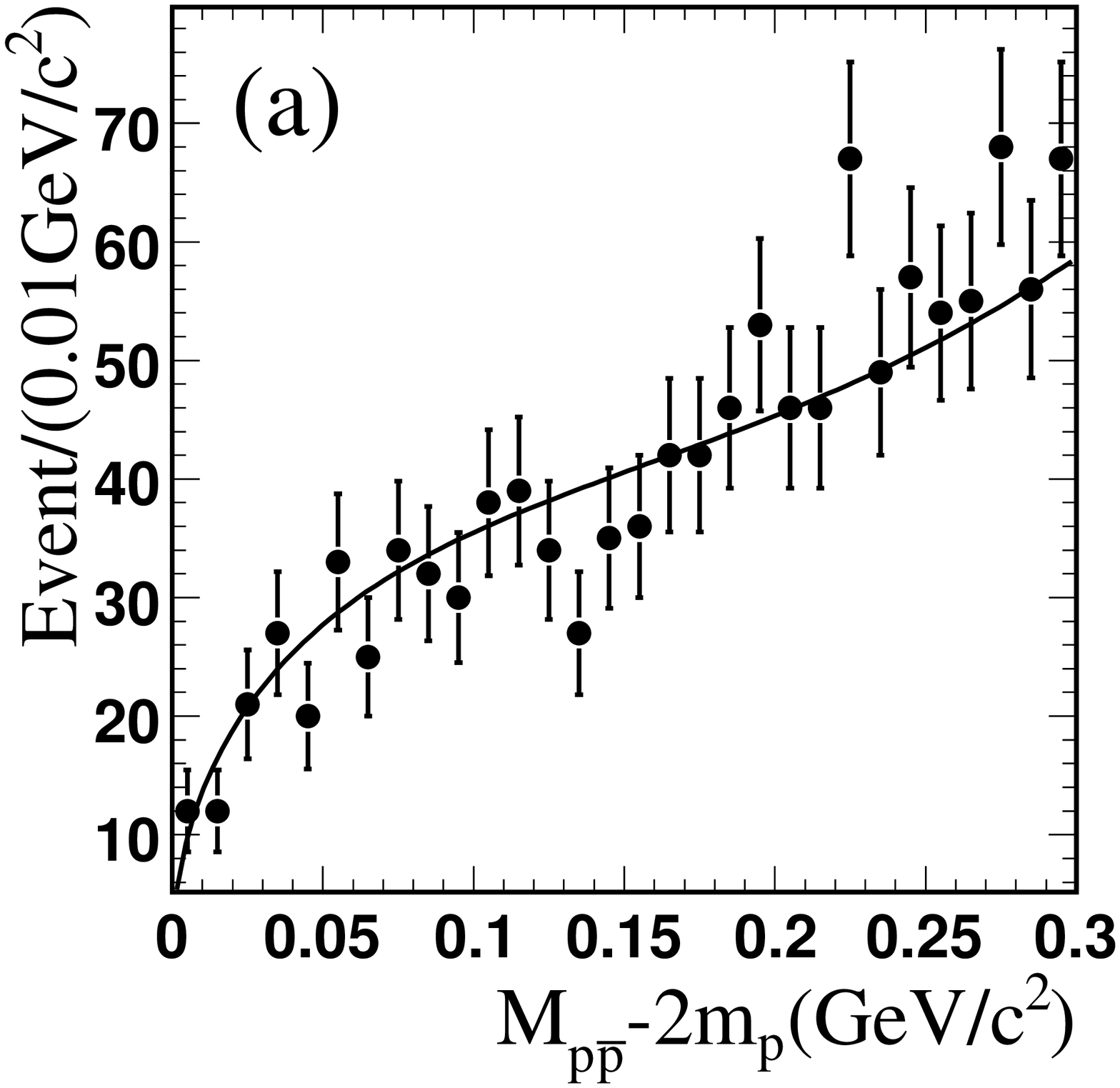}
\includegraphics[width=1.68in]{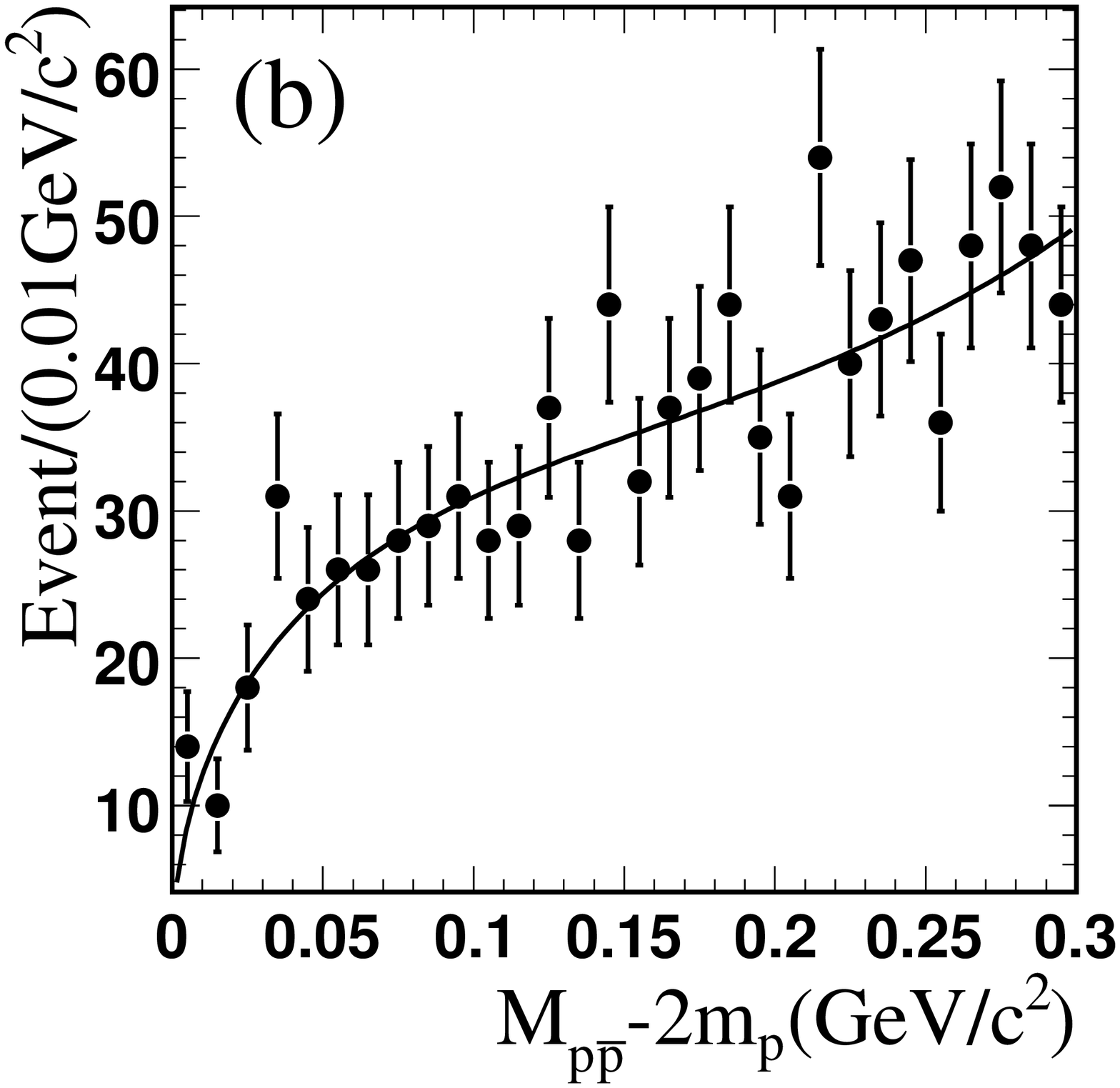}
\caption{\label{Fig2} The $p\bar{p}$ mass spectrum near threshold
for: $(a)$ selected $\psi^\prime\rightarrow\pi^{+}\pi^{-}J/\psi
(J/\psi\rightarrow\pi^0 p\bar{p})$ events for the same real data
sample. $(b)$ phase-space MC
$\psi^\prime\rightarrow\pi^{+}\pi^{-}J/\psi
(J/\psi\rightarrow\gamma p\bar{p})$ events that satisfy the
$\gamma p\bar{p}$ selection criteria. The smooth curves are the
results of the fit described in the text.}
\end{figure}

To characterize the $p\bar{p}$  threshold mass enhancement,
we  fit it  with an acceptance weighted
Breit-Wigner (BW) function of the form
$BW(M)\propto\frac{q^{2L+1}k^3}{(M^2-M_0^2)^2+M_0^2\Gamma^2}$,
where $\Gamma$ is a constant (determined from fit), $q$ is the
proton momentum in the $p\bar{p}$ rest-frame, $L$ is the $p\bar{p}$
orbital angular momentum, and $k$ is the photon momentum,
together with the function $f_{bkg}(\delta)$
with free normalization and
constants $a_1$ and $a_2$ fixed
at the $\pi^0 p \bar{p}$ phase-space MC values
({\it i.e.} the curves shown in Fig.~\ref{Fig2}) to represent the background
from mis-reconstructed $\pi^0 p\bar{p}$ events and a possible non-resonant
$p\bar{p}$ phase-space contribution. The BW is multiplied by
the MC-determined signal acceptance that is corrected for
MC and data differences  of the low momentum $\pi^+$ and $\pi^-$ tracking
efficiencies.  The tracking efficiencies determined from data
are measured  using  samples of tagged protons and antiprotons from the
process $J/\psi\rightarrow p\bar{p}\pi^{+}\pi^{-}$.

\begin{figure}
\includegraphics[width=2.4in]{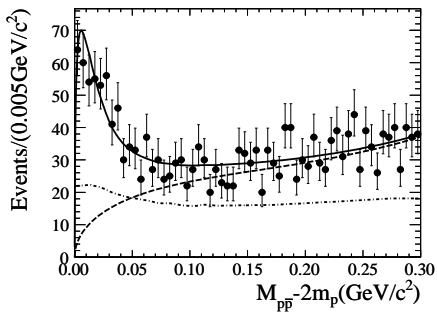}
\caption{\label{Fig3} The $p\bar{p}$ invariant mass spectrum for
the $\psi^\prime\rightarrow\pi^{+}\pi^{-}J/\psi
(J/\psi\rightarrow\gamma p\bar{p})$ after final event selection.
The solid curve is the fit result; the dashed curve shows the
fitted background function, and the dash-dotted curve indicates
how the acceptance varies with $p\bar{p}$ invariant mass.}
\end{figure}

The result of a fit using $L=0$ and confined to the
$M_{p\bar{p}}-2m_p<0.3~{\rm GeV}/c^2$ mass region is shown in
Fig.\ref{Fig3}. The fit returns a signal yield of
$519^{+36}_{-39}~{\rm (stat)}$, a peak mass of
$M=1861^{+6}_{-13}~{\rm (stat)}~{\rm MeV/}c^2$
 and a width of $\Gamma=0\pm23~{\rm MeV}/c^2$.
The fit quality is $\chi^{2}/d.o.f. = 42.6/56$.

In the above-described fit, the phase-space contribution is
treated as an incoherent background under the enhancement.
Possible fitting biases near threshold are investigated using a
set of MC samples that combine the signal with a uniform
incoherent phase-space background. In each MC sample, the mass,
width, and number of signal events are obtained from a fit using
the same procedure as that done on the data.  The averaged
differences between the fitted output and  input values are taken
as a systematic uncertainty associated with a possible fitting
bias. The r.m.s. of each parameter's bias measurements is taken as
the statistical error. The systematic uncertainties found from
varying the bin size and the fitting range are also included. The
total systematic error on the mass is $^{+7}_{-26}~{\rm MeV}/c^2$.
Including the systematic error, the upper limit on the width is
$\Gamma<38~{\rm MeV}/c^2$ at a $90\%$ confidence level.

We also tried to fit the $p\bar{p}$ mass spectrum  using known
particle resonances to represent the low-mass peak. There are two
spin-zero resonances listed in the PDG tables in this mass region:
the $\eta(1760)$ with $M_{\eta(1760)}=1756 \pm 9$~{\rm
MeV}/$c^{2}$ and $\Gamma_{\eta(1760)}=96 \pm 70$~{\rm
MeV}/$c^{2}$, and the $\pi(1800)$ with $M_{\pi(1800)}=1816 \pm
14$~{\rm MeV}/$c^{2}$ and $\Gamma_{\pi(1800)}=208 \pm 12$~{\rm
MeV}/$c^{2}$. A fit with $f_{bkg}$ and an acceptance-weighted
$S$-wave BW function with mass and width fixed at the PDG values
for the $\eta(1760)$ produces $\chi^2/d.o.f. = 144.8/56$; and
using the $\pi(1800)$ parameters produces $\chi^2/d.o.f. =
161.7/56$.

In summary, an anomalous strong, near-threshold enhancement in the
$p\bar{p}$ invariant mass distribution is observed in the decay
process of $\psi^\prime\rightarrow\pi^{+}\pi^{-}J/\psi
(J/\psi\rightarrow\gamma p\bar{p})$. If it is fitted with an
$S$-wave Breit-Wigner resonance function,
 the peak mass is
$M=1861^{+6}_{-13}~{\rm (stat)}^{+7}_{-26}~{\rm (syst)}~{\rm
MeV/}c^2$
 and the width is $\Gamma<38~{\rm MeV/}c^2$ at
the $90\%$ confidence level.  These values are consistent with the
published BESII results~\cite{ppb_jixb}. As indicated in
Ref.~\cite{x1835}, the ~$p\bar{p}$ mass threshold enhancement may
also be fitted with a broad structure $(\Gamma\sim100~{\rm
MeV}/c^2)$ multiplied by an FSI factor in Ref.~\cite{theory6}.
More precise measurement of the shape of the $p\bar{p}$ mass
threshold enhancement and more sophisticated fits such as
including some model dependent FSI factor in the fit will be
performed with much higher statistics $J/\psi$ data sample
collected at BESIII.

\vs 5mm

The BESIII collaboration thanks the staff of BEPCII and the
computing center for their hard efforts. This work is supported in
part by the Ministry of Science and Technology of China under
Contract No. 2009CB825200; National Natural Science Foundation of
China (NSFC) under Contracts Nos. 10625524, 10821063, 10825524,
10835001, 10935007; the Chinese Academy of Sciences (CAS)
Large-Scale Scientific Facility Program; CAS under Contracts Nos.
KJCX2-YW-N29, KJCX2-YW-N45; 100 Talents Program of CAS; Istituto
Nazionale di Fisica Nucleare, Italy; Russian Foundation for Basic
Research under Contracts Nos. 08-02-92221, 08-02-92200-NSFC-a;
Siberian Branch of Russian Academy of Science, joint project No 32
with CAS; the Chinese University of Hong Kong Focused Investment
Grant under Contract No. 3110031; U. S. Department of Energy under
Contracts Nos. DE-FG02-04ER41291, DE-FG02-91ER40682,
DE-FG02-94ER40823; WCU Program of National Reseach Foundation of
Korea under Contract No. R32-2008-000-10155-0. D. Cronin-Hennessy
thanks the A.P. Sloan Foundation.

\ed